# Assessing the Effectiveness of YARA Rules for Signature-Based Malware Detection and Classification

Adam Lockett
*Department of Computing and Informatics*
Bournemouth University
Bournemouth, United Kingdom
s5062305@bournemouth.ac.uk

*Abstract*— Malware often uses obfuscation techniques or is modified slightly to evade signature detection from antivirus software and malware analysis tools. Traditionally, to determine if a file is malicious and identify what type of malware a sample is, a cryptographic hash of a file is calculated. A more recent and flexible solution for malware detection is YARA, which enables the creation of rules to identify and classify malware based on a file's binary patterns. In this paper, the author will critically evaluate the effectiveness of YARA rules for signature-based detection and classification of malware in comparison to alternative methods, which include cryptographic and fuzzy hashing.

## I. Introduction

Signature-based detection is used by antivirus and malware analysis software to detect malware by comparing a repository of file signatures for known malware with signatures of files stored on a computer [1]. At the time of writing, most modern malware uses obfuscation and evasion techniques to avoid detection [2]. This reduces the effectiveness of traditional signature-based detection methods that use cryptographic hashing, as the malware has been modified so that its file signature does not match a known malware signature. Additionally, newly developed malware will not be detected, as its signature will not be included on malware signature databases until it has been identified by cybersecurity professionals. Therefore, alternative and more flexible approaches for signature-based detection are needed so that modified or obfuscated malware can be detected.

Alternative methods for signature-based detection include fuzzy hashing and YARA rules. In the next section, the author will define and address the strengths and limitations of these more recent approaches for detecting and classifying malware, in addition to traditional cryptographic hashing. In section III, the effectiveness of the different approaches will be evaluated by comparing malware detection and classification results of the approaches. Section IV will provide suggestions for improving YARA rules. Finally, section V will provide a conclusion regarding the effectiveness of YARA rules for signature-based malware detection, in addition to future work.

## II. Approaches for Signature-Based Malware Detection and Classification

### A. Cryptographic Hashing

Cryptographic hashing is a traditional method to detect and classify malware by generating a unique string (or hash) of hexadecimal characters for a file [3]. This enables known malware to be detected and classified on a computer, as file hashes generated are compared to a database of malware signatures, which includes information about the type of malware a signature relates to.

Although cryptographic hashing verifies if a file is identical to known malware, it will only detect exact matches. If a single bit is modified in a file, the hash value of the file will be different. Consequently, this means no similarities between files will be detected [4]. Therefore, cryptographic hashing is susceptible to malware that has been modified or obfuscated, as well as new types of malware. Moreover, this means that repositories of malware signatures need to be constantly updated.

### B. Fuzzy Hashing

Fuzzy hashing involves dividing a file into multiple blocks, where a hash value is calculated for each block, which are then concatenated to form a final hash value. When fuzzy hashes of files are compared, a similarity percentage is computed [5]. The hashing of multiple blocks of a file enables similarities to be detected between different files, as blocks that are identical from two different files will have the same block hash value.

Fuzzy hashing techniques also utilise distance algorithms to measure how similar different file blocks are. One fuzzy hashing technique, SSDEEP, uses the Damerau-Levenshtein distance function to determine the number of insertions, substitutions, deletions, and transpositions of characters or bits needed for different file blocks to match [6][7].

The use of fuzzy hashing techniques for signature-based malware detection provides greater flexibility in detecting malware, as an obfuscated or modified malware strain will likely be similar to the original strain of malware. This results in a high similarity percentage, which indicates a file is malicious. Conversely, an issue with fuzzy hashing is that it requires a repository of malware fuzzy hashes to determine if a file shares similarities with malware.

In comparison to cryptographic hashing, fuzzy hashing determines how similar a file is to other malware. Hence it is less susceptible to malware obfuscation techniques, whereas cryptographic hashing will only detect exact matches.

### C. YARA rules

YARA enables the creation of rules to identify and classify malware based on patterns of binary and textual data [8], where rules can utilise wildcards and regular expressions to detect these patterns [9]. Additionally, YARA rules provide a greater depth of malware information, as malware descriptions and metadata can be added to rules. Furthermore, YARA offers programmability, as rules can be created or adapted by anyone. Even so, creating effective YARA rules requires expertise in malware analysis, as weak YARA rules can result in malware evading detection [10].

YARA rules are also effective at detecting obfuscated malware, as many rules are built to detect different obfuscation capabilities, such as packing [11]. Moreover, rules have been created to detect anti-debugging capabilities, in addition to rules that identify the architecture of a file.

Overall, YARA rules offer more customisability and can supply more malware class information when compared to fuzzy hashing. However, fuzzy hashing is more flexible as it provides a percentage score on how similar a file is to malware. When compared to cryptographic hashing, both YARA rules and fuzzy hashing provide more flexibility and are more likely to detect new strains of malware, or malware that uses obfuscation techniques.

## III. EVALUATION OF APPROACHES

### A. Methodology

To assess the effectiveness of YARA rules for detecting and classifying malware, a controlled experiment was conducted. This involved applying YARA rules to malware samples to determine if a rule matches the sample and if matched rules gave a correct malware classification. Fuzzy and cryptographic hashes were generated for the same malware samples, with malware detection and classification results taken.

Both cryptographic and fuzzy hashing have many different hashing algorithms. For the experiment, SHA-256 was the hashing algorithm selected for cryptographic hashing, as it is a common hashing algorithm for identifying malware. SSDEEP is a fast, industry-standard algorithm that was selected for fuzzy hashing [12].

Fifteen malware samples were downloaded onto a VM from a malware repository on GitHub [13], as was a repository of YARA rules [11]. The malware samples downloaded from GitHub are well-known and all used some form of obfuscation. Another fifteen malware samples were downloaded from the VirusTotal malware repository for enterprise customers and students. These samples are relatively new and were uploaded to the repository in November 2021. The purpose of using two different sets of samples, one with older samples and one with recent samples, is to assess the effectiveness of the different approaches at detecting and classifying new malware that is not well-known.

Over 410,000 SSDEEP malware signatures from MalwareBazaar were used in the experiment for comparison with the malware samples. In the experiment, any matches with a similarity score of 50% or more were marked "detected as malicious".

VirusTotal was used to determine if cryptographic hashes generated from the GitHub malware samples were malicious and if they were classified correctly, whereas MalwareBazaar was used for the VirusTotal samples. MalwareBazaar was used instead because the malware samples will have already been fingerprinted by VirusTotal.

### B. Experiment Results

The following table presents the detection and classification results of the malware samples taken from the GitHub (GH) and VirusTotal (VT) repositories, using the approaches outlined in the methodology.

TABLE I. MALWARE SAMPLE DETECTION RESULTS

| Approach | Detected as Malicious | | Matches Malware Classification | |
|---|---|---|---|---|
| | GH | VT | GH | VT |
| SHA-256 Hash | 15 | 0 | 15 | 0 |
| SSDEEP Hash | 5 | 4 | 5 | 5 |
| YARA Rules | 12 | 5 | 6 | 2 |

From the quantitative results of the experiment shown in Table I, it is discernible that cryptographic hashing detects and classifies correctly all the malware samples from the GitHub repository. However, this is because SHA-256 hashes have already been generated for these malware samples, as they are known and have been analysed by threat hunters.

Conversely, none of the cryptographic hashes generated for the VirusTotal malware samples was found in MalwareBazaar. This is because the malware samples have only recently been analysed and uploaded (November 2021) by VirusTotal, which correctly identifies and classifies the malware. This suggests that cryptographic hashing is effective at detecting known malware, but not new or unknown malware.

Other studies also evaluate the effectiveness of cryptographic hashing. Results of a study by Sarantinos et al. were a 99.88% detection rate for cryptographic hashes generated by known malware samples [4]. This further suggests that cryptographic hashing is very effective at detecting known malware, but it will not detect unknown malware. Therefore, malware detection results for cryptographic hashing should not be compared with other approaches unless unknown malware samples are used [4]. Nevertheless, the results from the analysis of known malware using cryptographic hashing are still included to determine if the malware is classified correctly.

While the fuzzy hashing approach using SSDEEP detected and classified fewer malware samples than cryptographic hashing overall, it did provide multiple similarity matches with other known malware for many of the malware samples. The SSDEEP hash of one malware sample had two matches with malware signatures, with similarity scores of 100% and 97%. In addition to this, another malware sample had ten matches with a similarity score of 33%. This verifies that malware uses obfuscation and modification techniques to evade detection, thus fuzzy hashing approaches are useful for detecting obfuscated or new strains of a malware family, whereas cryptographic hashing cannot detect similarities between files unless they are identical.

In comparison to the fuzzy hashing results, the results for YARA rules are reasonably similar, although the results suggest that fuzzy hashing is more effective at classifying malware, whereas YARA rules are more effective at detecting malware and malicious capabilities. However, YARA rules are still effective at classifying malware, as rule matches can provide rich information about the capabilities and structure of the malware, which can be seen in the matched YARA rules for a malware sample in Figure 1.

```
-Keilhos:
SEH_Save
SEH_Init
escalate_priv
keylogger
win_registry
win_token
win_files_operation
Str_Win32_Winsock2_Library
maldoc_find_kernel32_base_method_1
IsPE32
IsWindowsGUI
IsPacked
HasOverlay
HasDebugData
HasRichSignature
```

Figure 1: YARA Rule Matches for Keilhos Malware Sample

Figure 1 shows the YARA rule matches for the Keilhos malware. Many of the rules, such as "keylogger", "maldoc_find_kernel32_base_method_1" and "IsPacked" indicate malicious activity, while also providing some classification and context regarding how it works. In this case, we might assume that the malware sample is a keylogger in the format of a malicious document that uses packing techniques to obfuscate itself.

Although some of the rules such as "SEH_Save" might only give context to malware analysts, once the YARA rules have been searched for on the internet there is usually more information about the purpose of the rule. For example, "SEH_Save" indicates the malware sample has anti-debugging capabilities [14]. This implies that YARA rules are effective at supplying classification information as well as potential tactics, techniques, and procedures (TTPs) of the malware.

The following table shows the number of malware samples that had matched at least one YARA rule or had a 1% or more SSDEEP similarity score. These results were included to show the context fuzzy hashing and YARA rules can supply for a malware sample, even if it is not classified or detected as malicious.

| Approach | Matches | |
|---|---|---|
| | GitHub | VirusTotal |
| SSDEEP Hash | 5 | 5 |
| YARA Rules | 14 | 8 |

TABLE II.    MALWARE SAMPLE MATCHES

From the results presented in Table II, both SSDEEP hashes and YARA rules had malware or rule matches that did not provide enough evidence for a malware sample to be detected as malicious. YARA rules had the most matches with a 73% match rate, which suggests that YARA rules are effective at providing some level of classification to obfuscated and new malware. Conversely, SSDEEP hashing only had a 33% match rate.

IV. SUGGESTIONS FOR IMPROVING YARA RULES

The results of the evaluation suggest that YARA rules are effective at detecting malware, but they did not correctly classify as many malware samples as the SSDEEP fuzzy hashing approach. Therefore, to increase the flexibility of YARA rules for both detecting and classifying malware, an improvement that could be made would be the functionality to use fuzzy hashing in YARA rules. This improvement has already been proposed by Naik et al. and would enable YARA rules to detect malicious capabilities based on the structural similarities of malware, which can't be done for standard YARA rules [15].

Another improvement that could be made to YARA rules would be more documentation, specifically the descriptions of popular rules for detecting and classifying malware. This is because descriptions were not always written for the rules used for the evaluation, which assists in classifying malware samples and providing context.

V. CONCLUSION

A. Overall Conclusion

In summary, from the experiment results and secondary research, it is evident that YARA rules are effective for signature-based detection and classification of obfuscated and new malware in comparison to cryptographic and fuzzy hashing. While cryptographic hashing is effective at detecting and classifying known malware, YARA rules can identify malicious capabilities in obfuscated malware that has not been fingerprinted and new malware strains, whereas cryptographic hashing can only detect known malware.

Although fuzzy hashing techniques can find similarities in files to detect obfuscated malware, it requires a repository of malware fuzzy hashes. Therefore, it may not recognise completely new types of malware. In contrast, YARA rules can detect malicious capabilities without a database of malware signatures.

From the results in Table I, YARA rules detect the most malware samples. This implies that YARA rules are the most effective signature-based technique for detecting malware, but it still only detects 57% of the malware samples. This emphasises the need for behavioural-based malware detection in addition to signature-based detection. However, considering that the malware samples were obfuscated or new strains, YARA rules are still an effective approach.

Overall, both cryptographic and fuzzy hashing are effective at detecting and classifying malware. However, YARA rules are the most effective for signature-based malware detection, as they can detect obfuscated and unknown malware, while providing enough context for it to give some level of classification.

*B. Future Work*

In this paper, we assessed the effectiveness of YARA rules for signature-based malware detection in comparison to cryptographic and fuzzy hashing. While these approaches are well-known, there are other alternative approaches for signature-based detection. One of these approaches is import hashing, where hash values are calculated based on a file's imported library functions and their order, which can be used to detect and classify malware [16]. In the future, the evaluation covered in this paper could be extended to include and assess import hashing for signature-based malware detection and classification.